\documentstyle[sprocl,epsf]{article}
\def\PLB{{\em Phys. Lett.}  B}
\def\PRL{\em Phys. Rev. Lett.\/}
\def\PRD{{\em Phys. Rev.} D}

\def\be{\begin{equation}}
\def\ee{\end{equation}}
\def\bea{\begin{eqnarray}}
\def\eea{\end{eqnarray}}

\def\lesssim{\le}
\def\gtrsim{\ge}

\begin{document}
\rightline{Report no: CLNS 97/1507}
\title{
NEGATIVE COUPLING INSTABILITY AND 
GRANDUNIFIED BARYOGENESIS}

\author{TOMISLAV PROKOPEC}
\address{Newman Laboratory of Nuclear Studies, 
         Cornell University, Ithaca, NY 14853-5001}


\maketitle
\abstracts{
In this talk 
I review my recent work with 
Brian Greene and Thomas Roos \cite{GreeneProkopecRoos}. First  
I discuss the effect of a negative cross-coupling 
on the inflaton decay in two scalar field theories.
Our main finding is a new effect, the 
{\it negative coupling instability,}
which leads to explosive particle production and 
very fast inflaton decay. 
Then I discuss the consequences of this instability 
for grandunified baryogenesis models. The novel aspect of 
this review is an intuitive explanation 
of the {\it negative coupling instability,} using 
field trajectories in the configuration space.
}

\section{Introduction}

The problem of the inflaton decay and the subsequent 
Universe reheat has received a lot of attention in 
recent years. 
At this conference an 
extensive discussion has been devoted to this problem
(see talks by D. Boyanovsky and S. Yu. Khlebnikov)
\footnote{Daniel Boyanovsky presented a new inflationary
scenario in which the classical limit gives an accurate
description of the problem. Sergei Khlebnikov argued that 
generically, in theories with weak couplings, the 
classical limit is reached when
"occupation numbers" of the field modes become large.}. 

A very brief history of the problem of the inflaton decay
 is as follows.
In 1990 Traschen and Brandenberger 
\cite{TraschenBrandenberger} discovered that 
the inflaton, when it starts oscillating after  
inflation, decays exponentially fast via 
parametric resonance. The inflaton decays in a 
few dozens oscillations, which is typically many orders of 
magnitude faster than the perturbative decay rate
\cite{old theory}. An example is the inflaton
decay via a quartic coupling $g\phi^2\chi^2/2$
to a second scalar field $\chi$. The tree-level 
decay rate is simply 
\begin{equation}
\Gamma(\phi\phi\rightarrow \chi\chi)=
\frac{g^2\Phi^2}{8\pi\omega_\phi}\,,
\label{eq: perturbative rate}
\end{equation}
where $\Phi$ is the inflaton amplitude, and 
$\omega_\phi$ its frequency. 
Since typically $g\Phi\ll \omega_\phi$, 
the decay time $\tau_{\rm decay}\sim\Gamma^{-1} $ 
is many orders of magnitude greater than
the expansion time 
$t_H\sim 1/H\sim 1/\omega_\phi$. 

\section{Parametric resonance in a toy model: 
the Mathieu equation}

Consider the simple two field case with the potential:
\begin{equation}
V(\phi,\chi) = \frac{1}{2} m^2_\phi\phi^{2} 
+\frac{g}{2}\chi^2\phi^2 \,.
\label{eq: V Mathieu}
\end{equation}
In a static universe the linearized mode equations for
$\chi$ can be written as the Mathieu equation 
\begin{eqnarray}
\frac{d^2\chi_k}{dz^2} &+& 
\bigl [A(k)-2q\cos(2 z)\bigr]\chi_k=0\,,
\label{eq: mathieu}\\
A(k) &=& \frac{\omega_\chi(k)^2}{(\omega_\phi)^2}
+2q\,,\qquad
q=\frac{g\Phi^2}{4(\omega_\phi)^2}\,,
\label{eq: A,q}
\end{eqnarray}
where $z=\omega_\phi t$, and $\omega_\chi(k)^2=k^2+m^2_\chi$
is the frequency squared of $\chi_k$.  
The instability chart of the Mathieu 
equation is depicted in figure~\ref{fig: 1}. The 
$\mu=0$ curves divide the chart into stable and unstable 
regions. For $g>0$, $A\ge 2|q|$, and $\mu_{\rm max} \le 0.3$,
as can be seen in figure~\ref{fig: 1}.
One can define $\mu$ as follows: the amplitude of the 
corresponding mode grows in one oscillation of
the inflaton $\phi$ by a factor $\exp 2\pi\mu$. 
Since the mode occupation numbers grow as the amplitude 
squared, this growth can be interpreted as particle 
production, and is termed {\it parametric resonance.}
On the other hand for $g<0$, $A\ge -2|q|$, and one can show
\cite{GreeneProkopecRoos} 
that $\mu\le 4|q|^{1/2}/\pi$, which can be $\gg 1$ when
$|q|\gg 1$. This phenomenon is what we mean by 
the {\it negative coupling instability.}
The question is can this instability occur in 
a natural physical setting?

\section{Parametric resonance in a realistic model}

To understand this question, consider the general 
renormalizable two scalar field potential with 
a ${\cal Z}_2\times{\cal Z}_2$ global symmetry
\begin{equation}
V(\phi,\chi) = \frac{1}{2} m^2_\phi\phi^{2} + 
\frac{1}{2}m^2_\chi\chi^{2}
+\frac{\lambda_{\phi}}{4}\phi^{4} +
\frac{\lambda_{\chi}}{4}\chi^{4} + \frac{g}{2}
\phi^{2}\chi^{2}
\label{eq: V all}
\end{equation}
in an expanding universe. The results I discuss in this section 
are obtained numerically in \cite{GreeneProkopecRoos}.
Assume a natural hierarchy of couplings 
(1) $\lambda_\chi>g>\lambda_\phi$ (when $g>0$), and 
(2) $\lambda_\chi>|g|>\lambda_\phi$, 
$r\equiv \lambda_\phi\lambda_\chi/g^2>1$  (when $g<0$).
In the former case the inflaton $\phi$ rolls down
in inflation and then
oscillates along the $\chi=0$ direction, leading to 
the standard growth of the resonant modes 
(in the $\chi$ direction). As a consequence, 
the inflaton amplitude $\Phi$ 
decays exponentially fast into the $\chi$ fluctuations
with  the decay time of order 
\begin{equation}
\tau_{\rm decay}\simeq \frac{1}{2\mu\omega_\phi}
\ln\frac{n_{\rm max}}{n_0}\,,\qquad \mu\lesssim 0.3
\end{equation} 
where $n_{\rm max}\sim 1/g$ and $n_0\simeq 1/2$ 
are the maximum and 
initial occupation numbers, respectively, and $\omega_\phi$ 
is the frequency of $\phi$ at the end of inflation.

When $g<0$, the {\it valley\/} of the potential (\ref{eq: V all})
is not any more along the $\chi=0$, but along
\begin{equation}
\tilde\chi_0^2 =  \cases{{-m_\chi^2-g\phi_0^2 \over\lambda_\chi} 
            & \hbox{for} $\quad m_\chi^2+g\phi_0^2<0$ \cr 
0 & \hbox{otherwise}} \,.
\label{eq: chi_0}
\end{equation}
If $\chi_0\approx\tilde\chi_0$ initially, then it remains so 
during inflation (as long as $|q|\gg 1/10$). More generally,
the valley (\ref{eq: chi_0}) is the attractor for the zero mode.
However, during inflation the superhorizon modes grow
(to generate seeds for structure formation, just like in 
ordinary inflationary theories).
On figure~\ref{fig: 2} we show the valleys of the potential 
both for $m_\chi^2=0$ and  $m_\chi^2>0$. In the former case 
the valley trajectories are defined by
$\chi_0(t)/\phi_0(t)\equiv
\tan\Theta=\pm (-g/\lambda_\chi)^{1/2}$, and one can easily show
that along the valley directions the initial stages of the
evolution look just like the positive $g$ case with 
$q_{\rm eff}=2|q|$ and $\mu_{\rm eff}\lesssim 0.3$. Indeed,
we observe this behavior in our numerical simulations. 
However, at certain point the backreaction effects from 
created particles become important: the $\chi$ excitations
become massive with $m_{\chi\rm \; eff}^2=m_\chi^2
+3\lambda_\chi\langle\chi^2\rangle
+g\langle\phi^2\rangle$, and the picture changes dramatically. 
The valley equation obeys (\ref{eq: chi_0}) so that for
$m_{\chi\rm \; eff}^2=|g|\phi(t)^2$ the valley has 
a characteristic curved shape as seen in figure~\ref{fig: 2}.
If at this time $\phi(t)$ is growing, and $\chi$ has
a small oscillating amplitude whose phase matches
onto the decaying solution, then $\{\phi(t),\chi(t)\}$ does
not follow the valley in (\ref{eq: chi_0})
but instead it climbs the ridge.
Consequently, the infrared $\chi$ modes "see" an inverted 
harmonic oscillator and their amplitude grows exponentially 
fast. This is what we mean by the {\it negative coupling 
instability.} In this case the mode amplitudes and 
field variances grow typically much 
faster than in the case of parametric resonance. Indeed,
in our numerical simulations we observe the instability
exponent $\mu\lesssim |q|^{1/2}$ for $|q|\gg 1$.
This instability occurs generically when
$\Phi^2\ge m_{\chi \;\rm eff}^2\ge |q|^{1/2}\omega_\phi^2$,
and hence even in the massless $\chi$ case due to the
backreaction effects.
Note the stochastic nature of the instability. 
In order to predict how each of the modes behaves, one
must be able to compute the phase of
$\chi_0$ at $|g|\phi^2(t)=m_{\chi\;\rm eff}^2$
very accurately. This sensitivity is illustrated in 
figure~\ref{fig: 3}, in which we show the dramatic change in
the variance  growth, when the $\chi$ mass
changes by a tiny amount. For example, when $|q|=350$, 
$\Delta m_\chi/m_\chi\simeq \pm 0.01$ leads to a change in 
the variance growth by a factor $\sim 10^3$. 
The instability peaks  
at $m_\chi^2=5.5\times 10^{-11}M_{\rm P}^2$ for which
$\mu_{\rm max}\simeq 20\simeq |q|^{1/2}$.

Note that a similar instability occurs when $m_\chi^2<0$,
$g>0$. In this case an investigation of the 
configuration space 
shown in figure~\ref{fig: 4}
indicates that the instability occurs
quite generically. Indeed, if at the end of inflation
$\phi_0^2\gg -m_\chi^2/g$ and $\chi$ is sufficiently
close to the valley value at $\chi_0=0$, 
once $\phi_0(t)^2< -m_\chi^2/g$ the infrared 
$\chi$ modes become unstable, leading to rapid
mode growth and particle production.  
A detailed study of this effect is under 
investigation. 

Without going into details, we now summarize 
the consequences of the negative coupling  
instability:
(1) the inflaton decays much faster, typically within 
a few oscillations (with $\langle\mu\rangle\sim 0(1)$);
(2) as a consequence of the following relation 
$A\ge 2|q|+|q|^{1/2}$, 
which specifies when the resonance shuts down, 
the field variances grow larger than in the positive $g$ 
case by a factor $\sim 4|q|^{1/2}$, and hence have a much 
stronger symmetry restoring force;
(3) massive $\chi$ particles are produced much easier, 
which may be relevant for grandunified baryogenesis (see below),
and the unstable momenta are of a much broader range
($\Delta k^2\sim |g|\Phi^2-m_\chi^2$). 

\section{Baryogenesis and the negative coupling instability}

In this section I discuss how the negative coupling 
instability can be used to facilitate
grandunified baryogenesis. I first summarize some of the 
results concerning production of heavy particles, and
then, on an example of a simple toy model, I
discuss how this heavy particle production
can affect grandunified baryogenesis.

 For a realistic choice of couplings, which is in chaotic
inflationary models constrained by the COBE satellite 
measurements
to be $\lambda_\phi\approx 3\times 10^{-13}$ or $m_\phi\lesssim
2\times 10^{13}\hbox{GeV}$, one can produce massive particles 
with
$m_\chi\sim 10^{14}\hbox{GeV}$ (as typically 
required by GUT baryogenesis
models), provided $|g|\gtrsim 10^{-8}$ and $\lambda_\chi > 
3\times
10^{-4}$ (for stability), see figure~\ref{fig: 5},
which corresponds to the parameter space $10^{-6}\gtrsim
|g|\gtrsim 10^{-8}$, $1 > \lambda_\chi > 3\times 10^{-4}$, 
leaving plenty of opportunity for baryogenesis model building.
This is to be contrasted with the positive $g$ case for which
$g\gtrsim 10^{-3}$ is
required. Such a large value leads to an unpleasant fine tuning
problem since the small value of $\lambda_\phi$ needs to be 
protected against radiative corrections.

To obtain an estimate of the baryon asymmetry that could be 
produced during preheating we study a simple toy model
\cite{KolbLindeRiotto}. 
In short, the model can be described as follows.  
Initially a certain amount of energy is transferred from
the inflaton to a heavy GUT scalar field {\it via\/} 
nonperturbative mechanisms, an example of which is 
the negative coupling instability. The result is a
cold (far from equilibrium) fluid of massive particles (with a 
mass $M_\chi$), which we assume decays in a B and CP 
violating manner into
light degrees of freedom that instantly thermalize. 
This out-of-equilibrium CP violating decay is the crucial step
that leads to net baryon production. In order to treat
the problem analytically we further assume that at all times the
temperature $T$ of the light relativistic particles is 
$\ll M_\chi$. The main results of this investigation are 
expressions for the final baryon-to-entropy ratios $n_B/s$
for a massless and massive inflaton. Here we quote the result
for the massless inflaton decay
\begin{equation}
\frac{n_B}{s}=
\epsilon
\left(\frac{\rho_X^0}{g_*M_X^4}
\right)^{\frac{1}{4}}
\left(\frac{\Gamma_X}{H_0}
\right)^{\frac{3}{4}}
\,,\quad {\rm when}\quad 
\rho_\phi^0\Gamma_\chi,\rho_\phi^0\Gamma_\chi
<g_*M_\chi^4H_0\,,
\label{eq: baryon-to-entropy ratio}
\end{equation}
where $\epsilon$ an effective CP violation (the decay of 
each pair $X-\bar X$ produces $\epsilon$ baryons),
$\rho_\phi^0$ and $\rho_X^0$ are the initial energy 
densities of the $X$ and $\phi$ fluid, respectively, 
$g_*\sim 10^2-10^3$ is the number of relativistic degrees 
of freedom, $M_X$ is the mass of $X$,  
$\Gamma_\phi$, $\Gamma_X$ are the decay rates for $\phi$ 
and $X$, respectively, and $H_0$ is the expansion
rate. In addition we have assumed a ``natural'' hierarchy
of time scales $H_0\gg\Gamma_\chi\gg\Gamma_\phi$.  
The remarkable feature of this result is that the 
final baryon-to-entropy ratio does not depend on the decay
constants $\Gamma_X$ and $\Gamma_\phi$.
Why this is so can be understood as follows. 
As long as the temperature never rises close to $M_X$ 
so that the massive particles $X$ are never 
re-populated by scattering off the radiation fluid,
most of the massive particles decay out of equilibrium 
and create $\simeq \epsilon N_\chi^0$ baryons,
independent on the decay rate $\Gamma_\chi$. 
The entropy production, if dominated by the late
inflaton decay, is also independent on 
the decay rate $\Gamma_\phi$, so that the final 
baryon-to-entropy ratio is independent on the decay 
rates. We believe that this feature lends more credibility 
to our result.

According to (\ref{eq: baryon-to-entropy ratio})
the final baryon to entropy ratio is proportional to 
$\rho_X^0$.  Since the energy density
of the (non-relativistic) $X$ particles is proportional to their
variance ($\rho_X\approx M^2_X\langle(\delta X)^2\rangle$),
the spikes in figure~\ref{fig: 5} 
are directly imaged to spikes in 
baryon production. For a massless inflaton the final baryon to 
entropy
ratio is expressed in terms of the maximum variance 
\begin{equation}
\frac{n_B}{s}=
\epsilon g_*^{-\frac{1}{4}}
\left(\frac{4g^2}{\lambda_\phi}
\right)^{\frac{3}{4}}
\frac{\langle\left(\delta X^2\right)^2\rangle_0}{M_X^2}
\,,
\label{eq: baryon-to-entropy ratio II}
\end{equation}
From figure~\ref{fig: 5}
we see that the largest value for the variance
$\langle\left(\delta X^2\right)^2\rangle_0\sim 10^{-9}M_{\rm 
P}^2$,
where $M_{\rm P}\simeq 2.4\times 10^{18}$GeV is the reduced 
Planck
mass, leading to a maximum value for the 
baryon-to-entropy ratio $n_B/s\sim 10^{-3}\epsilon$ (we set
$g_*=100$). We conclude that with our toy model
it is not hard to obtain baryon-to-entropy ratio consistent
with observation $n_B/s\sim 2-12\times 10^{-11}$.
Certainly more realistic models
for GUT baryogenesis at preheating will have to be constructed to
verify the viability of this scenario, but the preliminary 
results
are very promising.

\section*{Acknowledgements} 

This research was conducted using the resources of
the Cornell Theory Center, which receives major funding from the
National Science Foundation (NSF) and New York State, with 
additional
support from the Advanced Research Projects Agency (ARPA), the
National Center for Research Resources at the National 
Institutes of
Health (NIH), IBM Corporation, and other members of the center's
Corporate Partnership Program. 
Writing this review would not be possible without the
contribution of Brian Greene and Thomas Roos. I acknowledge 
NSF funding.

\section*{Figures}


\noindent 1. The stability chart for the Mathieu equation.
The dark regions are stable. In the unstable (light) regions we 
marked contours of constant instability index $\mu$. The contours 
shown are $\mu=0,0.1,0.2,0.3,0.5,1,2,3$. The lines $A=\pm 2|q|$
are also plotted.
\label{fig: 1}

\medskip
\noindent 2. The valleys of the potential (\ref{eq: V all}).
In the massless case oscillations are along  
$\chi_0(t)/\phi_0(t)=\tan\Theta=
\pm (-g/\lambda_\chi)^{1/2}$,
while in the massive case the trajectories are given by
(\ref{eq: chi_0}).
\label{fig: 2}

\medskip

\noindent 3. The $\chi$ field variances for three runs with slightly
different $\chi$ masses ($m^2_\phi=7.2\times 10^{-13}M^2_{\rm P}$,
$\lambda_\phi=10^{-12}$, $\lambda_\chi=10^{-5}$, $g=-10^{-9}$)
\label{fig: 3}

\medskip\noindent 4. The valleys of the potential with 
negative $m_\chi^2$ and positive $g$.
In the massless case oscillations are along  
$\chi_0=0$, while in the massive case the 
trajectories are given by
(\ref{eq: chi_0}).
\label{fig: 4}

\medskip
\noindent 5. Maximum peak and valley variances as a function
of $|g|$ for $m_\chi=10^{14}$GeV. The inflaton is massless
($m_\phi=0$), $\lambda_\phi=3\times 10^{-13}$, $g<0$ and
 $\lambda_\chi$ is adjusted to keep $r=10$.
\label{fig: 5}

\pagestyle{empty}

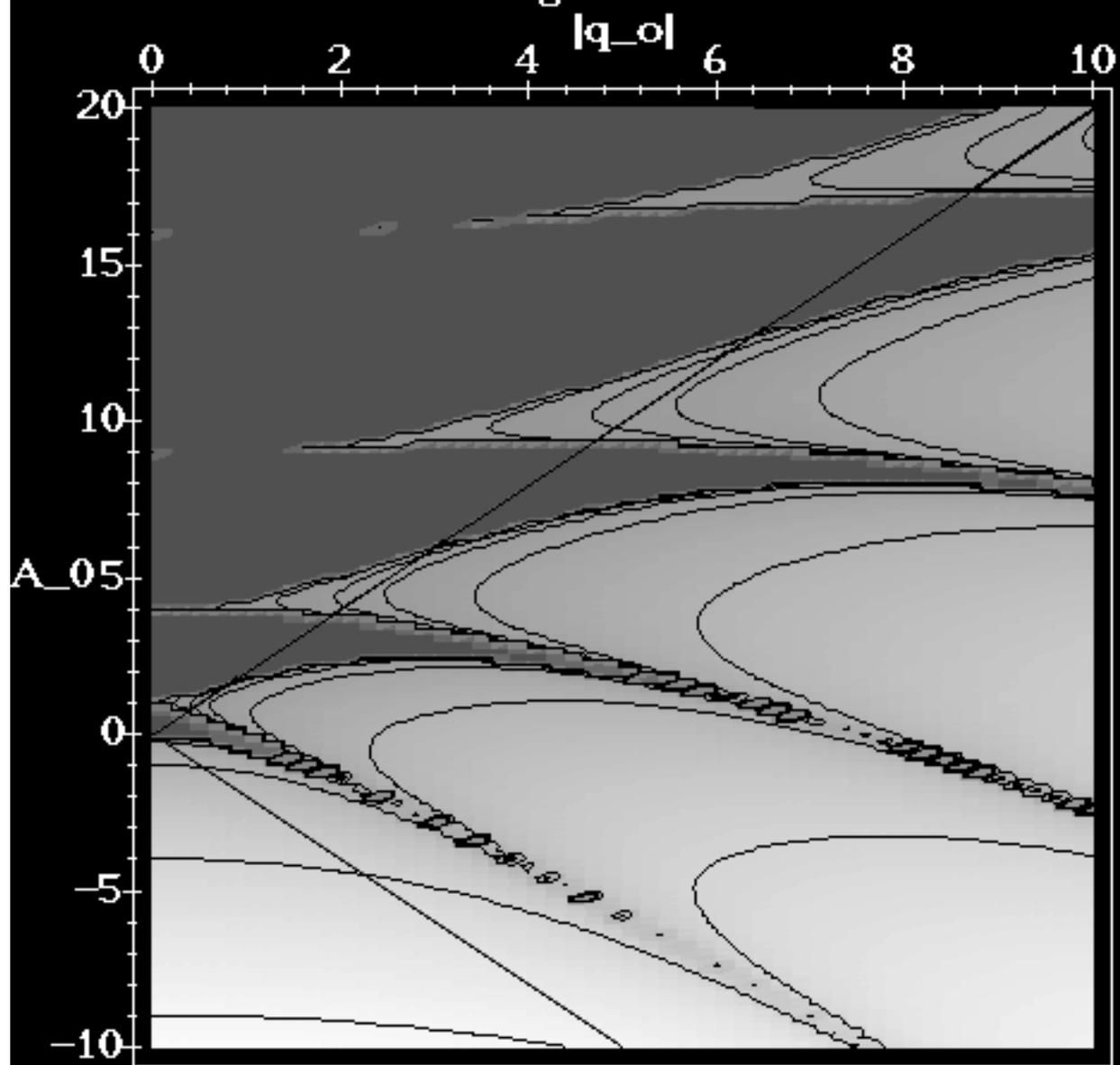
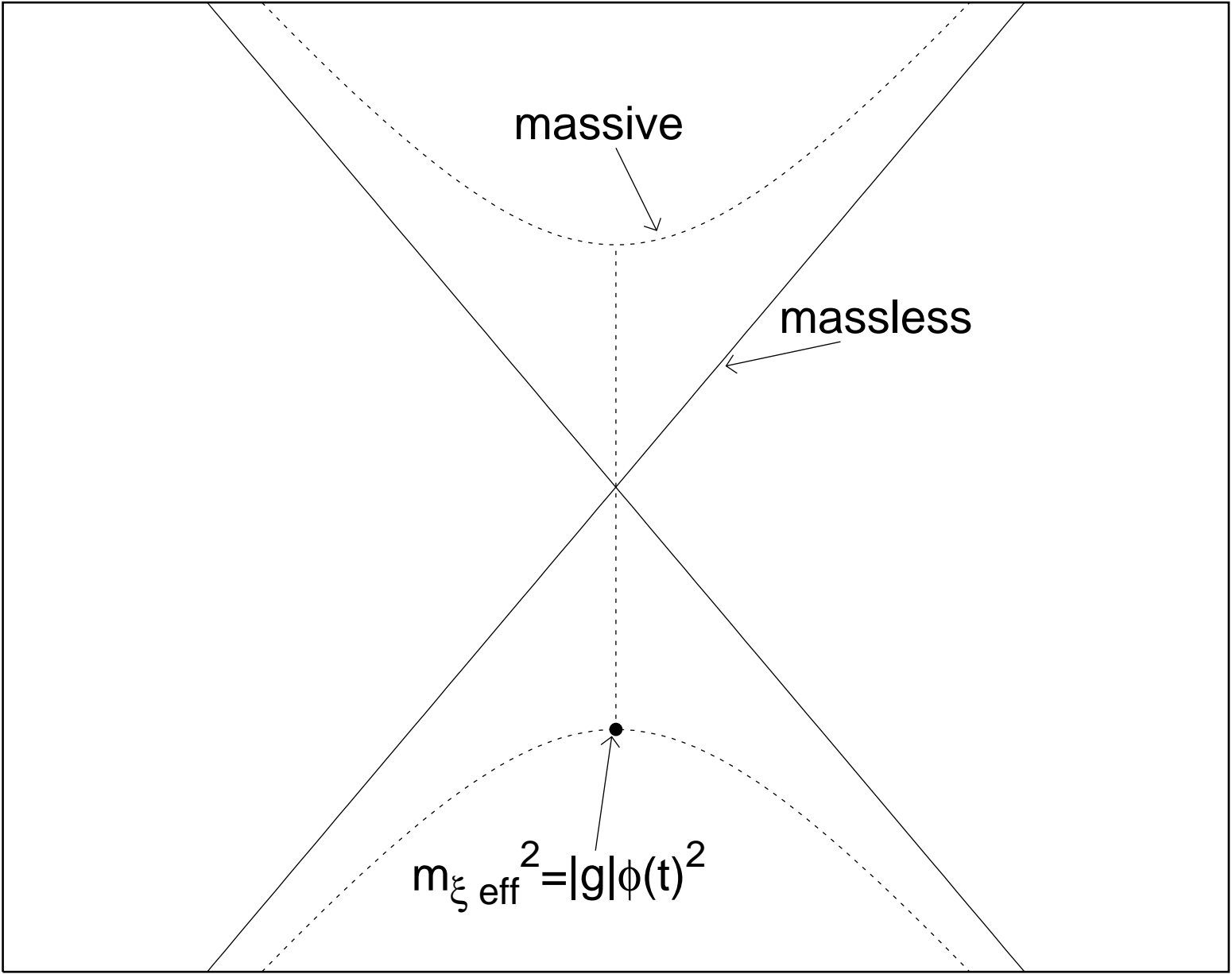
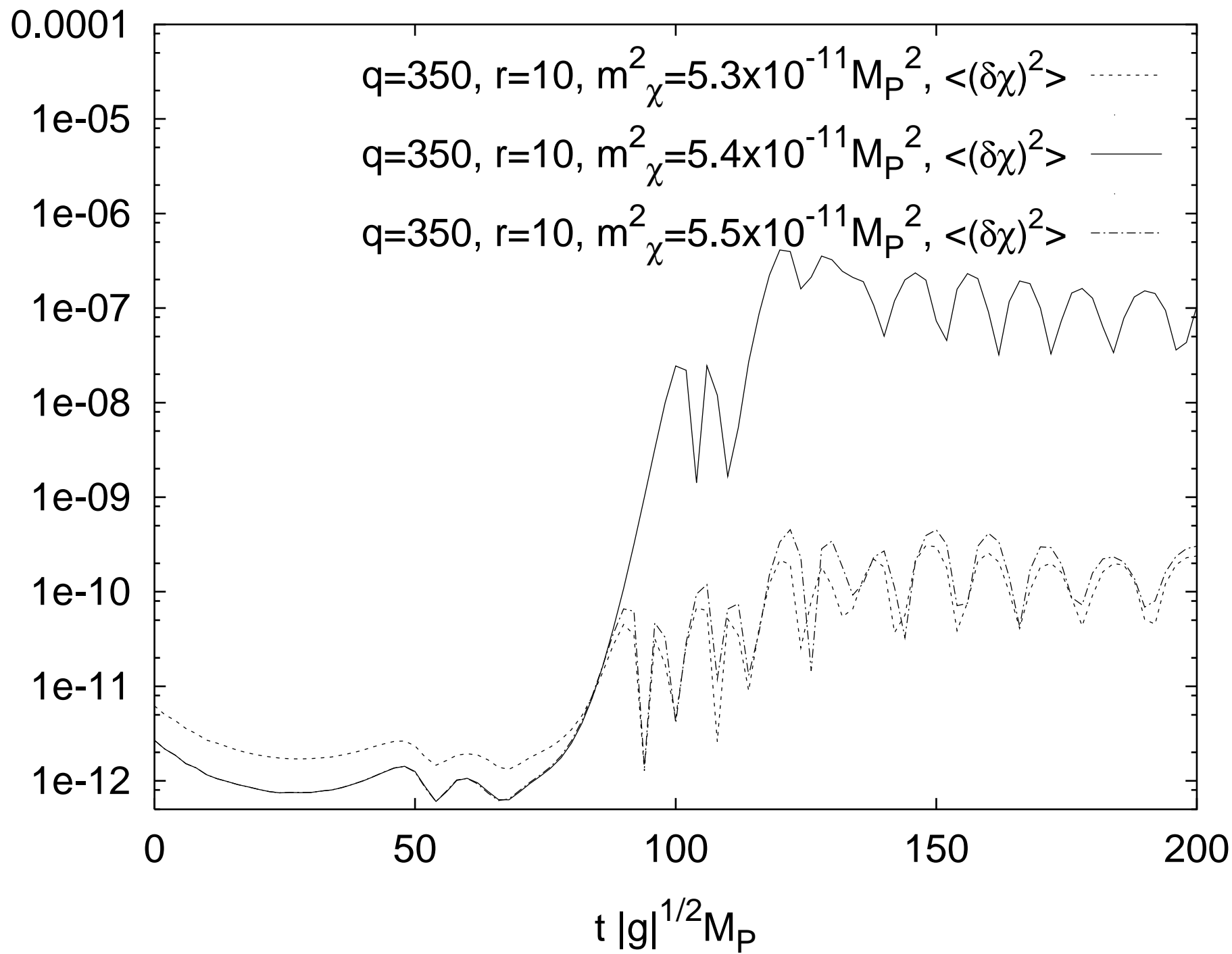
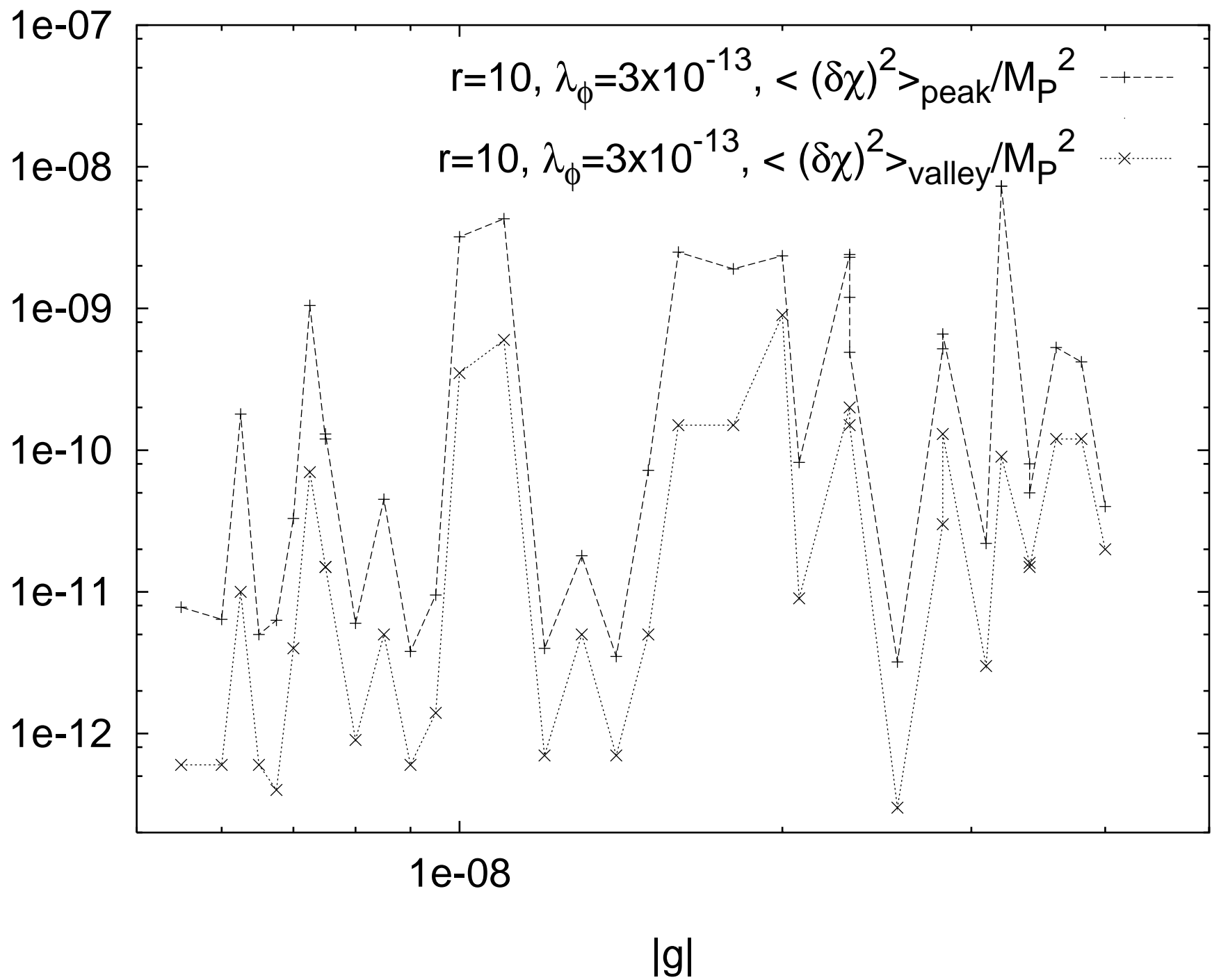
\end{document}